\documentclass[aps,prd,twocolumn,superscriptaddress,floatfix]{revtex4-1}
\usepackage{color}
\usepackage{amssymb}
\usepackage[normalem]{ulem}
\usepackage{graphicx,color}
\usepackage{amsmath}
\usepackage{subfigure}
\usepackage[normalem]{ulem}
\usepackage{booktabs}
\usepackage{xcolor}
\usepackage{epsfig,graphics,color,graphicx,amsmath}

\colorlet{darkgreen}{green!50!black}
\colorlet{brightyellow}{yellow!75!red}
\colorlet{orange}{red!50!yellow}
\colorlet{darkblue}{blue!60!black}
\colorlet{darkred}{red!80!black}

\newcommand{\sla}{\not\!}

\usepackage{soul}

\usepackage{mathtools}
\usepackage{mathrsfs}
\usepackage{bm}
\usepackage{relsize}
\usepackage{indentfirst}

\usepackage{xcolor}

 \usepackage{verbatim} 

\usepackage{amssymb,amsmath,multirow,epsfig,graphicx,color}

\usepackage{epsfig}
\usepackage{graphicx,amsmath}
\def\be{\begin{eqnarray} &&}

	\def\ee{\end{eqnarray}}

\def\sla#1{\rlap\slash #1}

\newcommand\ba{\begin{eqnarray}}
	\newcommand\ea{\end{eqnarray}}

\newcommand{\bas}{\begin{eqnarray*}}
	\newcommand{\eas}{\end{eqnarray*}}
\def\sla#1{\rlap\slash #1}

\newcommand{\bno}{\begin{eqnarray*}}
	\newcommand{\eno}{\end{eqnarray*}}

\def\sl
\usepackage{hyperref}
\usepackage{url}
\usepackage{hyperref}
\hypersetup{
	colorlinks=true,  %% false,  
	linkcolor=blue,   %% black,  %% red, blue,
	filecolor=magenta,      
	urlcolor=blue,    %% black,    %% cyan,
	citecolor=blue,   %% black,   %% Red   %% blue   %% black
} 
\begin{document}
	\title{Covariant form factors for spin-1 particles} 
	      %%%%  Exploring the covariant form factors for spin-1 particles} 
	\author{J.~P.~B.~C.~de Melo}
	%\email[E-mail: ]{joao.mello@cruzeirodosul.edu.br}
	\affiliation{Laborat\'orio de F\'\i sica Te\'orica e 
		Computacional - LFTC, 
		\\
		Universidade Cruzeiro do Sul and 
		Universidade Cidade de S\~ao Paulo (UNICID) 
		\\  01506-000 S\~ao Paulo, Brazil}
\date{\today}

\begin{abstract}
Spin-1 particles, is a 
fundamental bound state for the two quarks, and play a crucial role in elucidating the electromagnetic properties within the realm of hadronic physics. Their intrinsic relativistic nature mandates a quantum field theory (QFT) framework for a comprehensive analysis. In this investigation, we employ both the instant form of QFT and light-front quantum field theory (LFQFT) as our theoretical tools. While conventional LFQFT approaches predominantly focus on the plus component 
of the electromagnetic current to extract the properties of spin-1 vector particles, our study extends this analysis by systematically incorporating the minus component 
as well. Our findings demonstrate that achieving a rigorous equivalence between these distinct current components necessitates the inclusion of nonvalence terms within the electromagnetic current operator. This crucial inclusion serves to restore manifest covariance and ensures a robust consistency with the results independently
  derived from the more conventional instant form of quantum field theory.
\end{abstract}
\maketitle

%% \vspace{-0.5cm}
%% \begin{keyword}
%% vector particle, electromagnetic form factors, light-front,
%% zero-modes \end{keyword}
%% 
\section{Introduction} 
\label{introduction}
%%%
Quantum field theory provides a comprehensive framework for understanding the subatomic world, encompassing both strong and electroweak interactions. Specifically, quantum electrodynamics (QED) describes the electroweak interactions, 
while quantum chromodynamics (QCD) addresses the strong interactions. Together, these theories form the basis of our understanding of fundamental forces in nature. 
The primary aim of QCD is to elucidate bound states in terms of %%the 
fundamental constituents of matter: quarks and gluons. 

Various approaches rooted in QCD and 
quantum field theory exist for studying bound  hadronic states, encompassing both spin-zero 
 mesonic states, such as the pseudoscalar pion and kaon  mesons, 
 as well as spin-one states, including 	the rho, kaon star mesons, and the deuteron.	

 In this work, we employ light-front quantum field theory (LFQFT),
  also known as light-front dynamics, grounded in QCD, to investigate
   hadronic bound states. This approach is particularly 
beneficial due to the relatively simple kinematic boost properties~\cite{Terentev1976,Kondratyuk1980,Brodsky1998}. 
LFQFT serves as the foundational theory for this methodology,
 enabling us to integrate two critical aspects, 
 QCD and the constituent quark model.

 Quantum field theory on the light front presents 
 distinct advantages over traditional equal-time quantum 
 field theory. Nevertheless, certain physical processes exhibit 
 the breaking of rotational symmetry, which has
  prompted numerous studies proposing 
  that this symmetry breaking is attributed   the nonvalence contribuitions~\cite{Sales1998,deMelo1997}.

   	The   exploration of these nonvalence terms
  that must be included in the electromagnetic 
  current to ensure covariance in the light-front 
  approach, has been extensively examined in the literature.

  	The zero modes is fundamentally tied to the off-diagonal elements of the current operator matrix when expressed in the basis of Fock states. 
  	
  	Its influence on electromagnetic form factors becomes significant as the longitudinal momentum transfer, 
  	$q^+$, near zero. This influence manifests as a  nonzero contribution from these off-diagonal terms for the  nonvalence contribuitions for  the 
  	electromagnetic matrix elements of the 
  	electromagnetic current, which would otherwise be absent.
  	
  	In contrast, without the zero mode, the hadron form factors are reduced to a simple overlap integral of the initial and final light-front wave functions. This implies that the entire physical result is determined exclusively by the valence components, which are represented by the diagonal elements in the Fock state expansion.

\begin{widetext}
	%% \begin{multicols}{2}
	\begin{figure*}[!htb]
		\begin{center}
				\epsfig{file=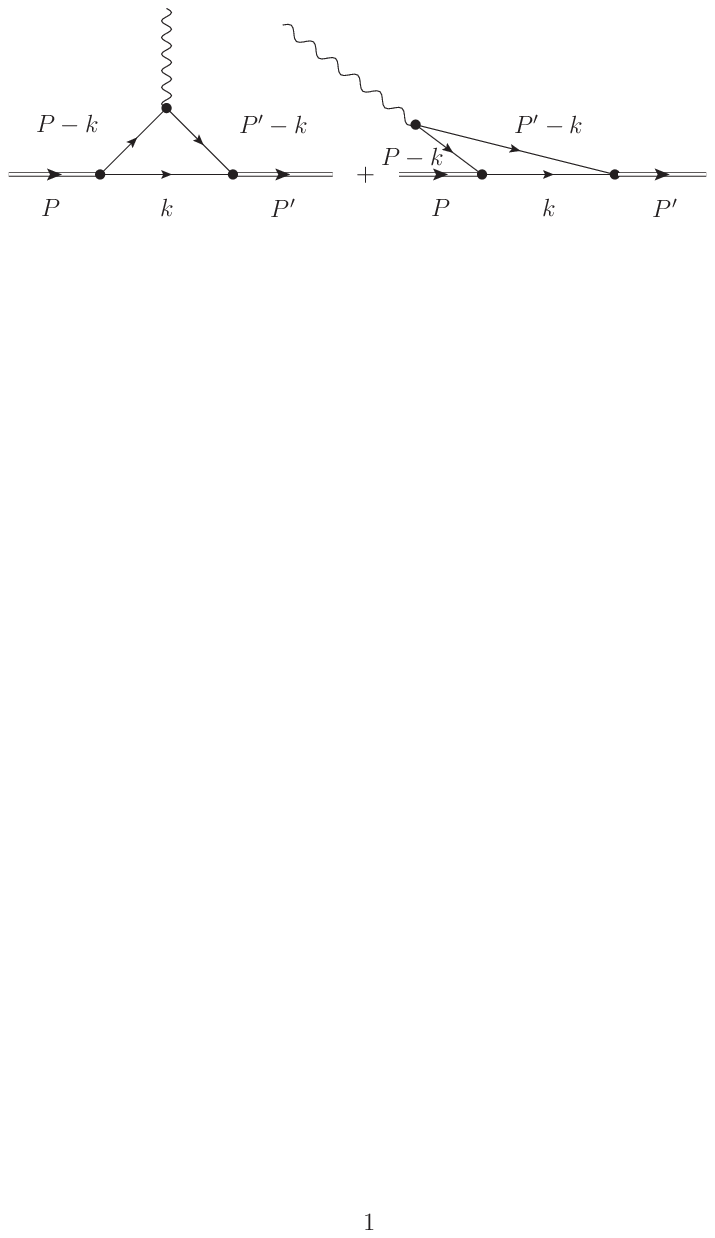, width=1.08\linewidth}
						\vspace{-18.cm}
			\caption{Feynman covariant triangle diagram, with the 
				valence and  nonvalence contribuitions to the 
				electromagnetic current.} 		   		
			\label{fig01}
	\end{center}  \end{figure*}
\end{widetext}

  	 In the light-front quark model, the full results for the
  	components of the electromagnetic matrix elements of the electromagnetic current, is the sum
  	of the valence components, and  nonvalence, which also contains the zero modes;  
  	 the Fig. \ref{fig01}, shows the valence and  nonvalence contributions to the electromagnetic current.

  The  research of the electromagnetic form factors, includes both
   scalar particles \cite{Naus1998,Sales1998,Lubo2018} and, also
   
    pseudoscalar mesons \cite{deMelo1999,Bakker2001,deMelo2002,Melikhov2002,deMelo1997,
  	daSilva2012,LeiPRL13,Yabusaki_2015,JVary2018,Choi2019,Moita2021,
  	Arifi2022,Jurandi2023a,Arifi2023v2,Jurandi2024},
 	 as well as vector

 	 particles \cite{Xu2019,
 	 	Frankfurt1988,Kobushkin1993,
 	 	Cardarelli1995,deMelo1997,deMelo1999gn,Lev1999,Jaus1999,
 	 	Bakker2002,Carlson2003,Jaus2003, deMelo2004cb,Choi2004,Dbeyssi2011,Lorce2009,deMelo2014fpa,
 	 	Mello2014b, Carrillo2015,
 	 	Krutov2016,Sun2016,Sun2017,deMelo2018,Meijian19,
 	 	Krutov2018,Lorce2022v1,2017APS,Liu2023,Pinto2024,Savtir2024,Acharyya2024,Parada2025}.

	In the case of pseudoscalar particles, such as the pion and kaon mesons, both
	 components of the electromagnetic current have been investigated to extract
	  observables, including electromagnetic form factors, electromagnetic radius, 
	  and electroweak decay constants~\cite{Moita2021}. 
	  This work aims to explore the calculation of the covariant electromagnetic
	   form factors $F_1$, $F_2$, and $F_3$, 
	as well as the related electromagnetic form factors of
	 charge $G_0(Q^2)$, magnetic \(G_1(Q^2)\), 
	and quadrupole \(G_2(Q^2)\). We will consider both the 
	plus and minus components of the electromagnetic current for
	 the rho meson scale, analyzing them at equal time and within
	  the light-front approach.
	
	It is important to note that calculating electromagnetic form factors using 
	different components of the electromagnetic current can sometimes lead to a
	 loss of covariance. In this study, we will investigate this loss of covariance
	  for spin-1 particles, focusing on the plus and minus components 
	  of the electromagnetic current.

To the best of our knowledge, the minus component of the electromagnetic current
 for spin-1 particles has not yet to be explored, especially regarding its 
 relationship to the covariant form factors

 The present work is organized as follows. 
 
 In Sec. \ref{introduction}, we provide 
 a overview of the problems addressed.   
 In Sec. \ref{current}, we formalize the expressions
  for calculating the covariant form factors in the case 
  of spin-1 particles and examine their relationships
   with the electromagnetic form factors. In 
   Sec. \ref{ffactors}, we calculate the elements of the electromagnetic 
   current based on the spin in the light-front framework,
    deriving these matrix elements in terms of the 
    Cartesian spin basis. We discuss the breakdown 
    of rotational invariance in relation to the angular 
    conditions at the light front. Finally, 
    in Sec. \ref{eleffactors},
     we compute the electromagnetic 
     factors using the model employed in this work and 
     present our results and conclusions.

%%II
 \section{Electromagnetic current and covariant form factors}
\label{current}

The most general expression for the spin-1 electromagnetic current, 
considering parity and time-reversal invariance, can be written as follows for the
 electromagnetic current
  (see Refs. \cite{Frankfurt1981,Frankfurt1993,deMelo1997,Carlson1997bs,Gilman2001} for details):

\begin{eqnarray}
%%  \begin{aligned}
J_{\lambda' \lambda}^{\mu} & = &  -\{ 
(p^{\prime \mu} + p^{\mu})
[F_1(q^2) (\epsilon_{\lambda'}. \epsilon_{\lambda})
\nonumber  \\ 
& - & \frac{ F_2(q^2)}{ 2 m^2_v} (q.\epsilon_{\lambda'} )
(q.\epsilon_{\lambda}) ] \nonumber \\
&  &  + F_3(q^2) 
((q. \epsilon_{\lambda'}) \epsilon_\lambda^\mu - (q.\epsilon_{\lambda}) 
\epsilon_{\lambda '}^{\mu} )
\}~.
\label{jmu}
%%  \end{aligned}
\end{eqnarray}

In the previous equation, 
$F_1$, $F_2$, and $F_3$ 
are the covariant electromagnetic form factors for spin-1 particles.
 Here, $m_v$ represents the vector bound state mass,
  while ($\lambda,\lambda^\prime$) 
  denote indices for the initial and final polarization 
  states, specifically~$(x, y, z)$.

The Lorentz covariant electromagnetic form factors, 
$F_1$, $F_2$, and $F_3$, are related to the electromagnetic form factors of 
	charge $G_C=G_0$, magnetic moment $G_M= G_1$, and 
	quadrupole $G_D=G_2$, 
	through the following relationships \cite{Gilman2001,Xu2019}:

	\begin{eqnarray}
	G_0 & = & F_1 + \frac{2}{3} \eta G_2,  \nonumber \\
	G_1 &= & F_3, \nonumber \\
	G_2 & = & F_1 -F_3 + (1 + \eta) F_2~.
	\end{eqnarray}

In the case of the zero momentum transfer, 
	these form factors are 	proportional to the  charge, magnetic 
	momentum and quadrupole momentum  \cite{Gilman2001,Xu2019}, 

	\begin{eqnarray}
	G_0(0)  & = & 1 \ \textrm{(units of e)}, \nonumber \\
	G_1(0) &  = & \mu_v  \ \textrm{(units of e/2 $m_v$)}, \nonumber \\ 
	G_2(0) & = & Q_v   \  \textrm{(units of e/$m^2_v$)}~.  
	\end{eqnarray}

 In the context of the light-front approach, we
  present the electromagnetic current for both components,  $(J^+=J^0+J^3, J^-=J^0-J^3)$,
   below. For further details, please refer to Refs.~\cite{deMelo1997,Frankfurt1981,Frankfurt1993}.

\begin{eqnarray}
%%  \begin{aligned}
J_{\lambda' \lambda}^{\pm} & = & - \{ (p^{'\pm} + p^{\pm})
[F_1(q^2) (\epsilon_{\lambda'}. \epsilon_{\lambda})
 \nonumber  \\ 
& - & \frac{ F_2(q^2)}{ 2 m^2_v} (q.\epsilon_{\lambda'} )
(q.\epsilon_{\lambda}) ] \nonumber \\
&  &  +  F_3(q^2) 
((q. \epsilon_{\lambda'}) \epsilon_\lambda^\pm - (q.\epsilon_{\lambda}) 
\epsilon_{\lambda '}^{\pm} )
\}~.
\label{j123}
%%  \end{aligned}
  \end{eqnarray}

The matrix elements for the electromagnetic current are calculated
 in the Breit frame, where the initial and final state momenta
  are 
  $p^\mu=\left(p_0,-q / 2,0,0\right)$ 
  and $p^{\prime \mu}=$ $\left(p_0, q / 2,0,0\right)$,
   respectively. The momentum transfer
    is $q^\mu=\left(p^{\prime \mu}-p^\mu\right)=(0, q, 0,0)$.
    
     The polarization vectors in the Cartesian basis are; 
initial state, $\epsilon_x^\mu=(-\sqrt{\eta},
 \sqrt{1+\eta}, 0,0), \epsilon_y^\mu=(0,0,1,0), \epsilon_z^\mu=(0,0,0,1)$,
  and, final state, 
  $\epsilon^{\prime \mu}=(\sqrt{\eta}, \sqrt{1+\eta}, 0,0), \epsilon_y^{\prime \mu}=(0,0,1,0), \epsilon_z^{\prime \mu}=(0,0,0,1),$
  
where $\eta=q^2 / 4 m_v^2$.

The light-front coordinates here, are defined by, the 
following quadrivector,~$a^\mu=(a^+=a^0+a^3,a^-=a^0 - a^3,(a_x,a_y)=a_\perp)$. 
The scalar product with the 
light-front approach, is given by the expression,  
$a^\mu b_\mu=\frac{1}{2}(a^+ b^- + a^- b^+) - \vec{a}_\perp \cdot  \vec{b}_\perp$,
and, the Dirac matrix at the light-front are defined as, $\gamma^+=\gamma^0+\gamma^3$, and, 
and $\gamma^-=\gamma^0-\gamma^3$~\cite{Brodsky1998}. 
The Dirac matrix, are related with the 
plus component of the electromagnetic current~$J^+_{ji}$, and minus component 
of the electromagnetic current~$J^-_{ji}$, respectively~\cite{deMelo1997}.  

For the plus components of the electromagnetic current, $J^+_{ji}$,
the following relations between the matrix elements of electromagnetic current, 
and the covariant form factors, are 
obtained below,

\begin{eqnarray} 
J^+_{xx} &= &  2 p^+\left[ ( 1 + 2 \eta) F_1 +   
\frac{q^2 ( 1 + \eta)}{2 m^2_v} F_2 \right]
\nonumber \\ & & 
-  ~2 q \sqrt{\eta} \sqrt{1+ \eta} F_3,
\nonumber \\ 
J^+_{yy} & = &  2 p^+ F_1,  \nonumber \\
J^+_{zz} & = &  2 p^+ F_1,  \nonumber \\
J^+_{zx} & = & - q  \sqrt{1+\eta} F_3,
\nonumber \\
J^+_{xz} & = & q \sqrt{1+\eta} F_3, \nonumber \\
J^+_{yx} & = & J^+_{xy} = J^+_{zy} = J^+_{yz} ~=~0 . 
\label{jijplus}
\end{eqnarray}

The covariant form factors, expressed
 as functions of the electromagnetic current matrix elements, 
 can be obtained for the plus component
  of the electromagnetic current using the matrix elements provided in Eq.(\ref{jijplus}), as, 
%% \begin{widetext}
\begin{eqnarray}
F_1 &  = &  \frac{J^+_{yy}}{2 p^+}  ~=~ 
 \frac{J^+_{zz}}{2 p^+},
 \nonumber \\ 
F_2  & = & \frac{m^2_v}{p^+ q^2 (1 + \eta)} 
\left[J^{+}_{xx} -  (1 + 2 \eta)  J^{+}_{yy}   - 2 \sqrt{\eta} J^{+}_{zx}    
\right],
\nonumber \\
F_3 & = & -\frac{J^+_{zx}}{q \sqrt{1+\eta}} ~=~ \frac{J^+_{xz}}{q \sqrt{1+\eta}}~.
 \label{f123plus}
\end{eqnarray}
%% \end{widetext}

Applying the same strategy to the electromagnetic current's
 minus component allows us to derive the following relations
  between this  component  and the covariant form factors, 

\begin{eqnarray}
J^-_{xx} &= & 2 p^- \left[(1+ 2 \eta) F_1 +
 \frac{ q^2 ( 1 + \eta)}{2 m^2_v} F_2\right]
 \nonumber \\
 & - &  2 q \sqrt{\eta} \sqrt{1+ \eta} F_3,
\nonumber \\ 
J^-_{yy} & = &  2 p^- F_1,  \nonumber \\
J^-_{zz} & = & 2 p^- F_1,  \nonumber \\
J^-_{zx} & = &  q \sqrt{1+\eta} F_3,
\nonumber \\
J^-_{xz} & = & - q \sqrt{1+\eta} F_3, \nonumber \\
J^-_{yx} & = & J^-_{xy} = J^-_{zy} = J^-_{yz} ~=~0 . 
\label{jij}
\end{eqnarray}
%% \end{widetext}

Analogous to the procedure used for the plus component
of the electromagnetic current, the covariant form factors,
$F_1,F_2$, and $F_3$
, are expressed below in terms of the minus component of the 
electromagnetic current:

%% \begin{widetext}
\begin{eqnarray}
F_1 &  = & \frac{J^-_{yy}}{2 p^-}  ~=~  \frac{J^-_{zz}}{2 p^-},
 \nonumber \\ 
F_2  & = & \frac{m^2_v}{p^{-} q^2 (1+\eta)} 
\left[J^+_{xx} - (1+ 2 \eta) J^{-}_{yy}  +  2 \sqrt{\eta}   J^{-}_{zx} 
\right],
\nonumber \\
F_3 & = & \frac{J^-_{zx}}{q \sqrt{1+\eta}} ~=~ -\frac{J^-_{xz}}{q \sqrt{1+\eta}}~.
 \label{f123minus}
\end{eqnarray}
%%   \end{widetext}

In the Breit frame, under the Drell-Yan condition, 
 ($q^+ =0$), we have the identity,  
 $p^+=p^{\prime+}=p^-=p^{\prime-}=p^0$. 
 We will utilize this equality in the subsequent analysis. 
It is well established that covariant electromagnetic
form factors in instant-form (equal-time) quantum 
field theory are independent of the specific electromagnetic 
current component chosen for calculations. However, this is 
not guaranteed in the light-front approach. The inherent 
loss of covariance, especially rotational symmetry, 
in the light-front formalism can yield different outcomes 
\cite{deMelo1997,deMelo2012,deMelo2019,Bakker2002}.

%% III
\section{LIGHT-FRONT  ELECTROMAGNETIC CURRENT AND ANGULAR CONDITION}
 \label{ffactors}
In this section, we calculate the elements of the electromagnetic current using the light-front spin basis. 
We then express these elements in terms of the Cartesian spin basis. 
Utilizing the polarization vectors in the spherical basis, 
we have the following expressions in the spin instant form basis~\cite{deMelo1997,Frankfurt1993},

\begin{eqnarray}
	\epsilon^\mu_+ & = &  -
	 \left( \frac{\epsilon^\mu_x+\imath \epsilon^\mu_y  }{\sqrt{2}} \right),
	 \nonumber \\
	\epsilon^\mu_0 & = & \epsilon^\mu_z ,
	\nonumber \\
	\epsilon^\mu_- & = &   \left( \frac{\epsilon^\mu_x- \imath \epsilon^\mu_y  }{\sqrt{2}} \right) ~.
\end{eqnarray}

The matrix elements of the electromagnetic current, focusing on the plus ($+$) and minus ($-$) components as defined in Eq. (\ref{j123}), are expressed below within the instant form spin basis. These elements describe the interaction of the hadron with an electromagnetic field and are fundamental
 for understanding  its 
 electromagnetic structure~\cite{Frankfurt1993,deMelo1997}
\begin{eqnarray}
J_{ji}^{\pm}   =  \frac{1}{2}  
\left( \begin{array}{ccc} 
J_{xx}^{\pm}+J^{\pm}_{yy}   &  \sqrt{2} J^{\pm}_{zx}   &  J_{yy}^{\pm}-J^{\pm}_{xx} \\
-\sqrt{2} J^{\pm}_{zx}      &  2 J^{\pm}_{zz}           &  \sqrt{2} J^{\pm}_{zx}     \\
J^{\pm}_{yy}-J^{\pm}_{xx}   &    -\sqrt{2} J^{\pm}_{zx} &  J_{xx}^{\pm}+J^{\pm}_{yy}  \\
\end{array} \right),
\label{eq:inst}
\end{eqnarray}
with the spin projections in the order~$m=(+,0,-)$, for the initial,
and final state polarization. 

 In the case of the light-front, the matrix elements $I_{m'm}^{\pm}$,
  for the plus and minus components of the  
  electromagnetic current are~\cite{deMelo1997}, 
\begin{eqnarray}
I^{\pm}_{m'm}   =   
\left( \begin{array}{ccc} 
  I^{\pm}_{11}    &  I^{\pm}_{10}   &  I^{\pm}_{1-1}    \\
  -I^{\pm}_{10}   &  I^{\pm}_{00}   &  I^{\pm}_{10}      \\
  I^{\pm}_{1-1}   &  -I^{\pm}_{10}  &  I^{\pm}_{11}       \\
\end{array} \right)~, 
\label{eq:lf}
\end{eqnarray}
The relations between the matrix elements
of the plus component of the electromagnetic 
current, $I^{+}_{m'm}$ (with $(m', m) \in \{-1, 0, 1\} \times \{-1, 0, 1\}$), 
in the Cartesian and light-front spin helicity bases, 
are given by the Melosh rotation 
\cite{Melosh1974,Chung1988, deMelo1999,deMelo2019}.
They are presented below~\cite{deMelo1997,Frankfurt1993},
\begin{eqnarray}
& & I^{+}_{11} = \frac{J^{+}_{xx}+(1+\eta) J^{+}_{yy}-
	\eta J^{+}_{zz}-  2 \sqrt{ \eta} J^{+}_{zx}}{2 (1+\eta)},
\nonumber \\
& & I^{+}_{10} = \frac{\sqrt{2 \eta} J^{+}_{xx}+\sqrt{2 \eta} J^{+}_{zz}
	- \sqrt{2} (\eta-1) J^{+}_{zx}}{2(1+\eta)},
\nonumber \\
& & I^{+}_{1-1} = \frac{-J^{+}_{xx}+(1+\eta) J^{+}_{yy}+
	\eta J^{+}_{zz} +  2 \sqrt{\eta} J^{+}_{zx}}{2 (1+\eta)},
\nonumber \\
& & I^{+}_{00} = \frac{-\eta J^{+}_{xx}+J^{+}_{zz} - 2 \sqrt{\eta} J^{+}_{zx}}
{(1+\eta)}~,
\label{ifront1plus}
\end{eqnarray}
for the plus component of the electromagnetic current; and for the minus 
component of the electromagnetic  current, $I^{-}_{m' m}$, we have the 
following relations below, 
%% 
%% \begin{widetext}
\begin{eqnarray}
& & I^{-}_{11} = \frac{J^{-}_{xx}+(1+\eta) J^{-}_{yy}-
	\eta J^{-}_{zz}-  2 \sqrt{ \eta} J^{-}_{zx}}{2 (1+\eta)},
\nonumber \\
& & I^{-}_{10} = \frac{\sqrt{2 \eta} J^{-}_{xx}+\sqrt{2 \eta} J^{-}_{zz}
	- \sqrt{2} (\eta-1) J^{-}_{zx}}{2(1+\eta)},
\nonumber \\
& & I^{-}_{1-1} = \frac{-J^{-}_{xx}+(1+\eta) J^{-}_{yy}+
	\eta J^{-}_{zz} +  2 \sqrt{\eta} J^{-}_{zx}}{2 (1+\eta)},
\nonumber \\
&  & I^{-}_{00} = \frac{-\eta J^{-}_{xx}+J^{-}_{zz} - 2 \sqrt{\eta} J^{-}_{zx}}
{(1+\eta)}~.
\label{ifront1minus}
\end{eqnarray}
%%  \end{widetext}

From the two equations above, Eq.~(\ref{ifront1plus}) and Eq.~(\ref{ifront1minus}), we observe that the plus and minus components of the electromagnetic current for spin-one particles, when expressed in the light-front basis, exhibit essentially the same structure. However, these two equations yield identical results only when calculations are performed using the equal-time approach. In contrast, the light-front formalism introduces  nonvalence contributions, also known as zero-mode  contributions \cite{deMelo1997,deMelo1999v2,deMelo2019},
 in addition to the valence contributions
 to the electromagnetic current matrix elements 
 (As illustrated in Figs.~\ref{pfig1} and \ref{pfig2}; a comprehensive treatment of  nonvalence or zero modes can be found in the detailed analysis
  presented by de Melo et al.  \cite{deMelo1999v2}.

In order to compute the electromagnetic current for 
spin-1 particles with the impulse approximation,
  we use the Mandelstam formula~\cite{Mandelstam1955,deMelo1997}, 
   for the plus and minus 
components of the electromagnetic current,~$J^{\pm}=J^0 \pm J^3$,  below,

\begin{eqnarray}
J^{\pm}_{ji} & = & \imath 
 \int \frac{d^4k}{(2\pi)^4}
 \frac{ Tr\left[\Gamma\Gamma\right]^{\pm}_{ji}\,
 	\Lambda(k,p_f)\,\Lambda(k,p_i) }
{((p_i-k)^2 - m^2+\imath\epsilon)} 
 \nonumber \\ 
& \times & \frac{1}
{(k^2 - m^2+\imath \epsilon)((p_f-k)^2 - m^2+\imath \epsilon) }.
\label{jcurrent}
\end{eqnarray}
For the expression of the  electromagnetic current,~Eq.~(\ref{jcurrent}) above, 
we have the following Dirac trace in the numerator, 
\begin{eqnarray}
Tr\left[\Gamma\Gamma\right]^{\pm}_{ji}& = & 
Tr[
 \epsilon_j\cdot  \Gamma(k,p_f)
(\sla{p_f} -\sla{k} +m) \gamma^{\pm}   
  \nonumber \\ 
  &   & (\sla{p_i} - \sla{k}+m) \epsilon_i 
\cdot \Gamma(k,p_i)(\sla{k}+m) ]\-.
\label{trace}
\end{eqnarray}

The regularization function in Eq. (\ref{jcurrent})
 is given by $\Lambda(k,p) = N/[(p-k)^2-m^2_R+ i \epsilon]^2$, 
 which is chosen to make the loop integration finite
 ~\cite{deMelo1997}. 
 The "Tr" symbol in the equations above denotes the Dirac trace.

The covariant electromagnetic current is calculated
 from Eq. (\ref{jcurrent}) by performing the integration
  analytically in the complex plane for $k^0$. 
  The remaining integrations over $\vec{k}$ 
  are then performed numerically.

In the light-front approach, the first integration in
 Eq.(\ref{jcurrent}), with respect to $k^-$, 
 is performed analytically using Cauchy's theorem;
  the subsequent integrations are then done numerically.

	The vertex model for the spinor structure of the composite spin-one particle, 
	($m_v:q\bar{q}$),  comes from the  model proposed in the 
	Ref.~\cite{deMelo1997}.  In the present work, 
	we use a more simplified version of the full vertex,
	i.e, $\Gamma(k,p) = \gamma^\mu$, 
	to simplify the exploration of the minus component of the 
	electromagnetic current.
	
	The wave function is obtained by employing the regularization function,
$\Lambda(k,p)$, with the light-front coordinates and the vertex $\Gamma^\mu (k,p)$
(for a more detailed explanation,  see the Ref. \cite{deMelo1997}),
	%%% 
	\iffalse 	
	With the regularization function, 
	$\Lambda(k,p)$, write with light-front coordinates, and the vertex $\Gamma^\mu(k,p)$, 
	nós obtemos a função de onda  %% dada no centro de 	massa do sistema, 
	(see more detalhes in the Ref.\cite{deMelo1997}),
	\fi
	%%
	\begin{eqnarray} 
	\psi(x,\vec{k})  = 
	%% \hspace{-0.15cm} 
	\frac{\cal{N}}{(1-x)^2 (m^2_v-M^2_0)(m^2_v-M^2_R)}
	%%\nonumber \\ 
	%% & \times & 
	%% \hspace{-0.15cm}
	  \epsilon \gamma.
	  \label{wfunction}
	\end{eqnarray}
	
	Where, $M^2_0$, is the free mass operator below, 
	\begin{eqnarray*}
	M^2_0 = \frac{k^2_\perp + m^2}{x} + 
	\frac{(p-k)^2_\perp+m^2}{1-x}-p^2_\perp~,
	\end{eqnarray*}
	
	and the function $M^2_R$ is  given by,
	\begin{eqnarray*}
	M^2_R = \frac{k^2_\perp + m^2}{x} +
	 \frac{(p-k)^2_\perp+M_R^2}{1-x}-p^2_\perp.
	\end{eqnarray*}

In the equations above, $x=\frac{k^+}{p^+}$, and the constant 
${\cal N}$ [in the Eq. (\ref{wfunction})], 
is obtained by imposing that the electromagnetic 
 charge form factor is equal to 1.

In the light-front approach, the matrix elements of the electromagnetic current,
following the equation, called  the "angular condition" 
\cite{deMelo1997,deMelo1999v2,Grach1984,Karmanov1996,Cardarelli1995}. 
After the use of the relation between' the matrix elements of the 
 electromagnetic  current,~$I^+_{m'm}$~(light-front basis) and the
  matrix elements for the electromagnetic current 
  in the   Cartesian spin basis,~$J^+_{ji}$, we have the following expression, 
  for the plus component of the electromagnetic current,
  
  \begin{eqnarray}
  	\Delta^+(Q^2) & = &   
  	(1+2 \eta)
  	I^{+}_{11}+I^{+}_{1-1} - \sqrt{8 \eta}
  	I^{+}_{10} - I^{+}_{00}
  	\nonumber \\ & =& (1 + \eta ) (J^+_{yy} -J^+_{zz})  = 0~;
  	\label{angcondplus}
  \end{eqnarray}
  
   and also, for 
the minus component of the electromagnetic current below, 

   \begin{eqnarray}
	\Delta^-(Q^2) & = &   
	(1+2 \eta)
	I^{-}_{11}+I^{-}_{1-1} - \sqrt{8 \eta}
	I^{-}_{10} - I^{-}_{00} \nonumber \\
	& =& (1 + \eta ) (J^-_{yy} -J^-_{zz})  = 0~.
	\label{angcondminus}
	\end{eqnarray}

In Figs. \ref{pfig5},
we can verify, for the matrix elements of the electromagnetic current, for both
components of the  electromagnetic current, plus component,
 as well for minus component, 
the Eq.(\ref{angcondplus}), and Eq.(\ref{angcondminus}),
are satisfied immediately at equal time,with the condition, $J^{\pm}_{yy}=J^{\pm}_{zz}$, 
for all values of $Q^2$.

On the other hand,
because  of the  zero modes, or  nonvalence contributions, that condition is 
not satisfied in the light-front approach; 
but, after all contributions added to the matrix elements of the electromagnetic 
currents, the angular condition equations are 
respected~\cite{deMelo1997,deMelo1999gn,deMelo2012,deMelo2019}, 
as can be seen in the Figs.~\ref{pfig5}.

The calculation at equal time is used to integrate the Feynman
 amplitude in four dimensions. In the present work, we employ
  the Breit frame, specifically with the condition $q^+ = q^0 + q^3 = 0$. 
  The integration is performed analytically first over $k_0$
   using Cauchy's theorem, and subsequently numerically 
   over the remaining spatial dimensions. In contrast, 
   the light-front approach involves the initial 
   integration over the light-front energy $k^-$ 
   in the internal loop. The subsequent integrations over
    $k^+$ and the transverse momentum $\vec{k}_\perp$ 
    are then carried out numerically.

In the case of the covariant calculation, or
 equal-time calculation, the numerical results for the plus ($+$) 
 electromagnetic or minus ($-$) components of the electromagnetic 
 current produced exactly the same results for the respective polarizations, Eq.~(\ref{jcurrent}) (see Figs.~\ref{pfig1} and \ref{pfig2}); 
 and the covariant form factors yielded the same numerical 
 results, independent of which component of the electromagnetic 
 current is utilized (see Fig.~\ref{pfig3}). In principle, electromagnetic
  form factors are Lorentz invariant; however, for light-front calculations,
   the matrix elements of the current calculated with both the $J^+$ and $J^-$
    components of the electromagnetic current need additional contributions,
     the zero modes contribution (sometimes referred to in the literature
      as the  nonvalence contribution), in order to have full covariance respected~\cite{Naus1998,Sales1998,deMelo1999gn}.

In the present work, we employ a technique to calculate the  nonvalence contributions to the valence current, a method referred to as the "dislocation pole method" (for detailed information, see the Refs. \cite{Naus1998,Sales1998,deMelo1999,deMelo1999gn}). Subsequently, the matrix elements for the electromagnetic current, computed using either the plus or minus component of the electromagnetic current, 
as shown in Eq.~(\ref{j123}), within the light-front approach, yield precisely the same results for both components of the electromagnetic current.

For the angular condition equation,
 one can also use the equations for the covariant form factors, $F_1^{\pm}$, Eq.~\eqref{f123plus}, and  Eq.~\eqref{f123minus},

 \begin{eqnarray}
  J^{\pm}_{yy} & =  &   2 p^{\pm} F_1, \nonumber  \\
  J^{\pm}_{zz} & = &  2 p^{\pm} F_1.
   \label{equalf1}
 \end{eqnarray}

Using the relations above, Eq.~\eqref{equalf1}, in the equations for the angular condition, Eq.~\eqref{angcondplus}, and Eq.~\eqref{angcondminus}, we obtain the following results for both components of the electromagnetic current:

 	\begin{eqnarray}
	\Delta^{\pm}(Q^2) & = &  (1 + \eta ) (J^{\pm}_{yy} -J^{\pm}_{zz})
	\nonumber \\
	 & = & (1 +  \eta) (2 p^{\pm} F_1 -  2 p^{\pm} F_1)=0~.
	\label{ang2}
	\end{eqnarray}

It is interesting to note that, when we calculate the covariant factors, 
both using the plus component of the electromagnetic current, as well 
the minus component of the electromagnetic current, 
we obtain the same values for these covariant electromagnetic form factors
; in particular in the case of the form factor
$F_1$, which is directly linked to the angular condition 
in the light front formalism, 
leads this condition to be immediately satisfied for both components of 
the electromagnetic current (see Eq.(\ref{ang2}) and the Fig.~\ref{pfig5}, 
for the angular condition, with both components of the electromagnetic current).

But, with the light-front approach, the results are
very different, mostly,  because of the  breaking of the 
rotational symmetry, 
and the contributions coming from the  of  nonvalence terms,
also called zero modes~\cite{deMelo1999gn,deMelo1997,Bakker2002,deMelo2019}, 
both for the component $J^+_{ji}$, as well as for $J^{-}_{ji}$  component
of the electromagnetic current, which can be seen
in figures for the electromagnetic matrix elements of the
current for which polarization  vector 
(see the Figs. \ref{pfig1} and \ref{pfig2}). 

For the $J^-_{ji}$ component of the electromagnetic current, 
calculated in the light-front approach, we have a very large breaking in 
rotational symmetry for all  matrix elements of the
 electromagnetic current$J^{-}_{ji}$,  (see Figs. \ref{pfig1} and \ref{pfig2}).

The electromagnetic form factors, $F_1$, $F_2$, 
and $F_3$, exhibit a significant dependence on the 
matrix elements of the electromagnetic current, particularly
 when considering the minus ($J^-$) components of the 
 electromagnetic current. 
For the Dirac form factor, $F_1$, presented in the top left
 panel of Fig.~\ref{pfig3}, our analysis reveals interesting 
 insights. When $F_1$ is calculated using the $J^+_{yy}$ component
  of the electromagnetic current within the light-front approach, 
  we observe excellent agreement with the covariant formalism 
  evaluated at equal times. This consistency indicates that
   rotational symmetry is preserved in this specific scenario.
    However, the use of the $J^-_{yy}$ component within the
     light-front formalism (represented by the solid blue 
     line in the figure) leads to a substantial violation 
     of rotational invariance.

In order to restore rotational symmetry, we incorporate
 the  nonvalence contributions to this component of the
  electromagnetic current (as illustrated in 
  Figs. \ref{pfig1} and \ref{pfig2}). 
Following this inclusion,
   we achieve precisely the same results, 
evident in the aforementioned figure.

In the Refs. \cite{deMelo1999v2,deMelo2012,deMelo2019},
 the plus  nonvalence components ($J^{+Z}$) of the electromagnetic current, was identified at the Dirac trace level for the microscopic current.
   Specifically, the terms that can lead to a loss of covariance were pinpointed for this component.  
   It has been demonstrated in the cited references above, that the valence terms (or zero modes) can be isolated within light-front quantum field theory for the $J^+$ component of the electromagnetic current. This isolation is achieved in terms of Dirac traces, specifically at the level of each polarization component. Notably, the conclusions presented are entirely model-independent~\cite{deMelo2012,deMelo2019}.

   The investigation of such terms for the other components of 
   the electromagnetic current for spin-1 particle case is ongoing.

 In the upper right panel of Fig.~\ref{pfig3}, we calculate 
 the electromagnetic form factor with the component $J^{\pm}_{zz}$, of electromagnetic current. As can be seen in Fig.\ref{pfig2},
 this element of the electromagnetic current, when calculated in the light front formalism, leads to
 a loss of rotational invariance, for both components of the 
 electromagnetic current; when we use the $J^+$ components, as well as,
 the component $J^-$. 
 Inspecting the Fig. \ref{pfig2}, we can see that the component 
 $J^+_{zz}$  of the electromagnetic current, does not have a sharp break
 when compared to the calculation with the instantaneous form;  however,
 the electromagnetic current component $J^-_{zz}$ has a large break for the 
 rotational symmetry. 
 
 The calculation of the electromagnetic form factor $F_1$, 
calculated with the components $J^{\pm}_{zz}$ of the 
electromagnetic current in the light front formalism, 
only with the valence terms of the electromagnetic current of 
components $J^{\pm}_{zz}$, leads to a different result than 
calculation at equal times calculated with both components 
of the electromagnetic current, as we can 
see in Fig.~\ref{pfig3}, top right panel. 
Note that after we add the  nonvalence terms to the current elements, 
the calculation of the electromagnetic form factor $F_1$, 
 (the black and orange dashed curves correspond to the figure respectively), 
we have exactly the calculation made with the 
usual quantum field theory at equal time.

To emphasize the importance of considering not
 only valence components but also  nonvalence components in the matrix elements of the electromagnetic current when employing the light-front formalism, we present in Fig.~\ref{pfig4}, the calculation of the differences
for the electromagnetic form factors
$F_1$, $F_2$, and $F_3$. These form factors are calculated 
both at equal time, and on the light-front.
As the presented results demonstrate, neglecting the addition of  nonvalence 
terms leads to a violation of rotational symmetry, 
yielding significantly different results and thus a strong breaking of this symmetry.     
This rotational symmetry is restored upon the inclusion of these  nonvalence terms in
 the respective elements of the electromagnetic current, for the plus,
 and, also the minus component of the electromagnetic current.
  
 \section{Electromagnetic form factors and results}
\label{eleffactors}

The electromagnetic form factors of a spin-1 
hadronic bound state, 
for charge
$G_0$, magnetic $G_1$, and quadrupole $G_2$, 
 can be expressed through the matrix elements of 
 the electromagnetic current \cite{Carlson1997bs,deMelo1997,deMelo2012,Gilman2001}.
 Utilizing the light-front and instant form bases, and 
 following the prescription outlined in Ref.~\cite{Grach1984}, 
the plus ($+$) and minus ($-$) 
components of these matrix element are given by:

\begin{eqnarray}
G_0^{GK}& = &\frac{1}{3}[(3-2 \eta) I^{\pm}_{11}+ 2 \sqrt{2 \eta} I^{\pm}_{10} 
+  I^{\pm}_{1-1}]  \nonumber   \\
& = &  \frac{1}{3}[J_{xx}^{\pm} +(2 -\eta) J_{yy}^{\pm} 
+ \eta  J_{zz}^{\pm}], \nonumber \\
G_1^{GK} & = & 2 [I^{\pm}_{11}-\frac{1}{ \sqrt{2 \eta}} I^{\pm}_{10}]
=J_{yy}^{\pm} -  J_{zz}^{\pm} - \frac{J_{zx}^{\pm}}{\sqrt{\eta}},
\nonumber \\ 
G_2^{GK}&=&\frac{2 \sqrt{2}}{3}[
\sqrt{2 \eta} I^{\pm}_{10} - \eta I^{\pm}_{11}-  I^{\pm}_{1-1}] 
=  \nonumber \\
 & & 
 \frac{\sqrt{2}}{3}[J_{xx}^{\pm}-(1+\eta) J_{yy}^{\pm} 
+ \eta  J_{zz}^{\pm}]~.
\end{eqnarray}
As recently explored in Refs. \cite{deMelo2012,deMelo2019}, the adoption of the aforementioned formulation stems from the prescription detailed in \cite{Grach1984}. This prescription elegantly eliminates the $I^{+}_{00}$ component from the possible combinations of the four matrix elements of the electromagnetic current, denoted by $I^{+}_{m'm}$. This elimination is crucial because the $I^{+}_{00}$ component is responsible for the emergence of zero modes, which represent  nonvalence contributions to the electromagnetic current. Notably, the expressions derived for the electromagnetic form factors of spin-1 particles using this approach yield results that are in exact agreement with equal-time calculations when employing the
 plus component of the electromagnetic current \cite{deMelo2012,deMelo2019}.

However, as we can see in Fig.~\ref{pfig10}, even using 
the prescription by Inna Grach {\it et al.}, with 
the minus component of the electromagnetic current in the light-front approach,
 the results differ from the instantaneous form, that is,
we have zero modes or  nonvalence mode contributions for the calculation of the electromagnetic form factors,  in case we use the minus component of the electromagnetic current; in other words, the valence sector is not enough 
to extract the electromagnetic form factors  for spin-1 particles with the minus component of the electromagnetic current.  
When we add for the matrix elements the minus component of the electromagnetic current, these  nonvalence terms, and or zero modes, we restore the covariance, and it can be verified,
the both components of electromagnetic current produce the same results.

For spin-one particles, such as the deuteron 
\cite{Carlson1997bs,Gilman2001}, the charge electromagnetic form factor of the $\rho$-meson exhibits a zero, indicating a change in sign \cite{Cardarelli1995,deMelo1997,Roberts2011,Pitschmann2012}. In this work, this zero is found to be approximately $Q^2_{zero} \sim 3~\text{GeV}^2$ (as illustrated in Fig.~\ref{pfig10}), where the charge electromagnetic form factor vanishes, $G_0(Q^2_{zero}) = 0$. This analysis considers the plus and minus components of the electromagnetic current calculated within the instant form framework. 

Notably, as depicted in the figure for the
 charge electromagnetic form factor, $G_0(Q^2)$, the light-front calculation employing the plus component of the electromagnetic current and the prescription of 
Grach et al.~\cite{Grach1984,deMelo1997} yields numerically equivalent results to equal-time calculations. This equivalence arises because this specific prescription ensures that zero-modes, or  nonvalence contributions, do not contribute to the electromagnetic current \cite{deMelo2012,deMelo2019}.

Taking the universal ratios between the electromagnetic form factors, 
given in the Refs.~\cite{deMelo2019,Hiller1992,deMelo2016},
 we get an estimate for the 
zero of the electromagnetic charge form factor,
 $Q^2_{zero} = 6 m^2_\rho \sim 3.5$ $GeV^2$, 
for the experimental bound state mass of the rho meson, 
$m_\rho = 0.775$ $GeV$~\cite{PDG}; so that the
 value predicted in this work is close to this value.

For the calculation involving the minus 
component of the electromagnetic current, the zero
 of the charge electromagnetic form factor, $G_0(Q^2)$, 
 is observed to occur around $Q^2_0 \sim 1.5 \, \text{GeV}^2$.
 
  This observation is made when considering solely the valence component of the electromagnetic current in the form factor calculation
   (as depicted by the dashed red line in the left panel 
  of Fig. \ref{pfig10}).

After including  nonvalence terms (or zero-modes) to the  matrix 
elements of the electromagnetic current minus, 
we have the same value for this zero, obtained in the instantaneous form and, also with the plus component of the
electromagnetic current. In the light-front, that zero have some dependence with the prescription in the light-front approach is adopted \cite{deMelo1997,deMelo2019},
but, after the  zero modes  be added to the matrix elements of the electromagnetic current, for both, plus and minus
components of the electromagnetic current, the zero positions  are the 
same \cite{deMelo2019}. Remembering that in the present work, we used the 
prescription from Inna Grach \cite{Grach1984}, which was demonstrated in 
previous works,  be free from  pairs terms, or zero modes \cite{deMelo2012,deMelo2019} 
for the plus component of the electromagnetic current.

As in the case of the  charge  electromagnetic form factor, 
$G_0(Q^2)$, the magnetic form factor, $G_1(Q^2)$, 
both for the plus component of the electromagnetic current, as well as 
for its minus component, when calculated at equal times, 
we get the exact same results, as can be seen in the 
Fig. \ref{pfig10}, above, right panel. In the case of
 the light-front formalism, however, with the calculation of this electromagnetic form factor, using the minus component of the electromagnetic current, we have a very different results (dashed red line), when compared with the calculation for both at times equal, as on the light front, with the plus component of the current electromagnetic (the other curves in the figure, which are impossible to distinguish, as they provide the same numerical results). 
Anyway, after adding the  nonvalence terms (or zero modes) 
to the minus component of the electromagnetic current; 
we get exactly the same results with the minus component of the electromagnetic current, 
it is impossible to distinguish these curves, as these calculations 
produce the same numerical results (see the Fig.~\ref{pfig10}).

The quadrupole moment for the rho meson is obtained with 
the limit for $Q^2$  goes to zero~\cite{deMelo1997,Hiller1992};

and the quadrupole electromagnetic 
form factor being negative for other values of $Q^2$, which can be seen in Fig.~\ref{pfig10}.  As for the electromagnetic and charge form factors, calculations performed at the equal times, 
for both components of the electromagnetic current, give the same numerical 
results,   we can see in the figure, the black and blue curves,  are totally indistinguishable.
 On the other hand, the calculation on the light front also reproduces the same results with the prescription from the Ref. \cite{Grach1984}, with the plus component of the electromagnetic current  (orange line in the figure). 
However, when calculating this electromagnetic form factor using the minus component of  the electromagnetic current, taking into account only the valence matrix elements of the minus 
component of the current electromagnetic, we have a very significant difference in the  results (dashed red line in the figure). 
But, upon considering the  nonvalence terms (or zero modes) of the matrix elements of the minus component of the electromagnetic current, we observe that the solid green curve in Fig. $\ref{pfig10}$ yields identical results to those obtained using the plus component of the electromagnetic current, both at equal times and within the light-front formalism \cite{deMelo2012,deMelo2019}.

For the numerical calculations, the vector meson mass is $m_v = 0.775$ GeV. The constituent quark masses are $0.430$ GeV, and the regulator mass, $m_R = 3.0$ GeV, matches the value used in Ref.~\cite{deMelo2019}. This choice was made to reproduce the experimental properties of the $\rho$-meson, as detailed in Ref.~\cite{PDG}.

In summary, this work employs the plus ($J^+$) and minus ($J^-$) components of the electromagnetic current to investigate the covariant electromagnetic form factors of spin-1 particles. We also analyze the related electromagnetic properties: charge ($Q$), magnetic dipole moment ($\mu$), and electric quadrupole moment ($Q_{el}$). Our analysis reveals that within the light-front formalism, to maintain rotational invariance, it is essential to include  nonvalence (or zero-mode) contributions to the matrix elements of the electromagnetic current. This inclusion is necessary not only for the current components themselves but also for the resulting electromagnetic form factors, thereby restoring covariance. We are currently considering the use of the full vertex $\Gamma^\mu$ \cite{deMelo1997} for both the plus and minus components of the electromagnetic current, aiming to separately analyze the contribution of each component to the complete vertex structure.

\begin{center}
{\bf ACKNOWLEDGMENTS}
\end{center}

 I thank Douglas A. Antonio for his careful reading of the manuscript.
This work was supported in part by 
the Conselho Nacional de Desenvolvimento Cient\'{i}fico e Tecnol\'{o}gico~(CNPq), Brazil. 
 Funda\c{c}\~{a}o de Amparo \`{a} Pesquisa do Estado de S\~{a}o Paulo (FAPESP), Brazil, 
No.2023/09539-1, and was also part of the projects, 
Instituto Nacional de Ci\^{e}ncia e Tecnologia-Nuclear Physics and Applications~(INCT-FNA,MCTI), Brazil,
Process No. 408419/2024-5,
and CNPq, CAPES and FAPERJ; 
%% Process No.464898/2014-5, 
 and FAPESP, Brazil,  thematic project, No. 2024/17816-8.
 %% \end{widetext}

\begin{center}
{\bf DATA AVAILABILITY}
\end{center}
	
The data that support the findings of this article are not publicly available upon publication because it is not technically feasible and/or the cost of preparing, depositing, and hosting the data would be prohibitive within the terms of this research project. The data are available from the authors upon reasonable request.

  %% \clearpage
      \newpage
  
%% \begin{widetext}
%% \begin{multicols}{2}
%% FIG. 1a and 1b
 \begin{figure*}[thb]
 	\begin{center}
\epsfig{figure=prdfig1.eps,width=7.5cm,angle=0}
\vspace{0.5cm}
 \epsfig{figure=prdfig2.eps,width=7.5cm,angle=0} 
  \caption{Matrix elements of the electromagnetic current $J^{\pm}_{xx}$~(left panel).
 	Right panel: The matrix elements of the electromagnetic current $J^{\pm}_{yy}$.
 	The matrix elements calculation at equal time, for both components of the 
 	electromagnetic current, "$J^+$", or "$J^-$",  provide exactly the same results. 
 	This does not happen with the light-front calculations;
 	  but affter added the  nonvalence contribuitions, the final results are identical. In that figures, full Light-front, 
 	means the sum of two contribuitions, i.e., 
 	valence plus  nonvalence contribuitions to the 
 	matrix elemements of the current.
}
 \label{pfig1}
 \end{center} 
 \end{figure*}
%% \end{widetext}

\vspace{2.cm}

%% \clearpage  \newpage  
%% \begin{widetext}
%% \begin{multicols}{2}
%% FIG. 2a and 2b
 \begin{figure*}[thb]
 	\begin{center}
\epsfig{figure=prdfig3.eps,width=7.5cm,angle=0}
\vspace{0.5cm}
 \epsfig{figure=prdfig4.eps,width=7.5cm,angle=0} 
 \caption{Matrix elements of the electromagnetic current $J^{\pm}_{zx}$~(left panel).
Right panel: The matrix elements of the electromagnetic current $J^{\pm}_{zz}$.
The matrix elements calculation at equal time, for both components of the 
electromagnetic current, "$J^+$", ou "$J^-$", give exactly the same results, 
as in the calculation of the elements of electromagnetic current above.
%% 
%% For the Ligh-front calculation is not the case, but, afterwards 
%% added the zero-modes, or  nonvalence contribuitions, 
%% the final results are identical.  
%% 
In that figures like the figures above, 
the full Light-front, is the sum of two contribuitions, i.e., 
valence plus  nonvalence contribuitions to the 
matrix elemements of the current. 
}
 \label{pfig2}
 \end{center} 
 \end{figure*}
%% \end{widetext}
	
%%  vspace{-15.cm}
\begin{figure*}[thb]
	\begin{center}
		\epsfig{figure=prdfig8.eps,width=6.5cm,angle=0}
		%%  \vspace{0.5cm}
		\epsfig{figure=prdfig9.eps,width=6.5cm,angle=0} 
		\caption{Angular condition for the plus componente of the electromagnetic current 
			(left panel), and the 
			minus componente of the electromagnetic current (right panel). 	In the case of the covariant calculation, for both components of the electromagnetic current, Eqs. (17) and (18), the angular condition, results in
			zero (solid black line).
			%%in the Fig.(\ref{pfig5}). 
			In the light-front approach, the angular condition for both components of the electromagnetic current is not satisfied (solid red line); with the addition of  nonvalence terms to both components of the electromagnetic current, we obtain the angular condition, i.e., zero (solid green line).
		}
		\label{pfig5}
	\end{center} 
\end{figure*}
%% \end{widetext}
%%% \clearpage  \newpage

%% line 833
%% \begin{widetext}
%% \begin{multicols}{2}
%%   fig3  FIG. 1a and 1b
\begin{figure*}[thb]
	\begin{center}
		\epsfig{figure=prdfig5.eps,width=7.5cm,angle=0}
		\vspace{0.65cm}
			\epsfig{figure=prdfig5b.eps,width=7.35005cm,angle=0}
		\vspace{0.25cm} 
		\\
		\epsfig{figure=prdfig6.eps,width=7.5cm,angle=0}
			%% 	\vspace{1.80020cm}
		\epsfig{figure=prdfig7.eps,width=7.35005cm,angle=0} 
		\vspace{-0.20cm}
		\caption{Upper left panel:
			 The covariant electromagnetic form factor~$F_1(q^2)$, calculate with the 
			matrix elements of the electromagnetic current,~$J^{\pm}_{yy}$ for equal time, 
			and, also with the light-front approach.
			~Upper right panel: Electromagnetic form factor~$F_1(q^2)$, calculated with the electromagnetic matrix 
			elements~$J^{\pm}_{zz}$. 
			Below panel: $F_2(q^2)$ electromagnmetic covariant form factor; and also, $F_3(q^2)$ 
			electromagnetic covariant form factor~(right panel).
		}
		\label{pfig3}
	\end{center} 
\end{figure*}
%% \end{widetext}   %%% ff1zzv1.eps 
	
\newpage

%% line 1115
%% \begin{widetext
%%   fig4.fig4
\begin{figure*}[thb]
	\begin{center}
		\epsfig{figure=deltaf1.eps,width=6.5cm,angle=0}
		\vspace{0.65cm}
		\epsfig{figure=deltaf1zz.eps,width=6.35005cm,angle=0}
		\\
		\epsfig{figure=deltaf2.eps,width=6.5cm,angle=0}
		%%	\vspace{1.80020cm}
		\epsfig{figure=deltaf3.eps,width=6.35005cm,angle=0} 
	%% 	\vspace{-.20cm}
		\caption{Upper left panel: 
			The diferences for the covariant electromagnetic form factor
			$F_1(q^2)$, between the equal 
			time calculation and Light-front approach; calculate with the 
			matrix elements of the electromagnetic current,~$J^{-}_{yy}$ for equal time, 
			and, also with the light-front approach, without and with the 
			 nonvalence contribuition.
			~Upper right panel: The same label, but, now calculate with 
			the $J^-_{zz}$ matrix elements of the electromagnetic current.
			\newline
			Below panel: Same as above, for $F_2(q^2)$ and $F_3(q^2)$.
		}
		\label{pfig4}
	\end{center} 
\end{figure*}
%% \end{widetext}   %%% ff1zzv1.eps 

	\newpage 

 %% \begin{multicols}{2}
 \begin{figure*}[thb]
 	\begin{center}
 		\epsfig{figure=prdfig10.eps,width=6.5cm,angle=0}
 		\vspace{0.5cm}
 		\epsfig{figure=prdfig11.eps,width=6.5cm,angle=0}
 		\\
 		\vspace{1.20cm}
 		\epsfig{figure=prdfig12.eps,width=6.50cm,angle=0}
 		\caption{Upper left panel: 
 			The charge electromagnetic form factor~$G_C=$~$G_0(q^2)$, 
 			and, upper right panel, the magnetic form factor~$(G_M=G_1)$;
 			calculated with the plus and minus components of the electromagnetic current, 
 			for	both, in the instant form and light-front approach. Lower panel:  
 			the quadrupole form factor, ($G_D$=~$G_2(q^2)$), 
 			labels like the upper panels. 
 			\label{pfig10}
 		}
 		\end{center} 
 	\end{figure*}
 	%% \end{widetext} 

 \clearpage 
  \newpage 

\onecolumngrid

\bibliography{currentv2.bib}

%%eeeeeeeeee
\end{document}